\documentclass[prd,showpacs,preprintnumbers,nofootinbib,eqsecnum]{revtex4}

\usepackage[dvipdfm]{graphicx}
\usepackage{latexsym}
\usepackage{amsfonts}
\usepackage{amssymb}
\usepackage{amsmath}

\begin{document}

\preprint{YITP-12-1, KUNS-2378}

\title{Tunneling without barriers with gravity}
\author{Sugumi Kanno$^{1}$}
\author{Misao Sasaki$^2$}
\author{Jiro Soda$^{3}$}

\affiliation{$^{1}$Institute of Cosmology,
Department of Physics and Astronomy, 
Tufts University, Medford, Massachusetts 02155, USA \\
$^2$ Yukawa Institute for Theoretical Physics, Kyoto University,
Kyoto 606-8502, Japan\\
$^{3}$Department of Physics,  Kyoto University, Kyoto 606-8502, Japan\\
}%

\date{23 January, 2012}

\begin{abstract}

We consider the vacuum decay of the flat Minkowski space to an 
anti-de Sitter space.
We find a one-parameter family of potentials that
allow exact, analytical instanton solutions describing
tunneling without barriers in the presence of gravity. 
In the absence of gravity such instantons were found by Linde
and rediscovered and discussed by Lee and Weinberg more 
than a quarter of a century ago. 
The bounce action is also analytically computed.
We discuss possible implications of these new instantons to cosmology
in the context of the string theory landscape.
\end{abstract}

\maketitle
\section{Introduction}

Ever since the spontaneous symmetry breakdown of a vacuum
was found in field theory,
its implications to the history of our universe have been a matter
of great interest in cosmology and much work has been done on
how the vacuum decay proceeds dynamically.
One of the most important contributions to the understanding
of the vacuum decay was made by Coleman who developed a method 
to describe the vacuum decay by considering $O(4)$ symmetric 
instantons that mediate the vacuum decay~\cite{Coleman:1977py,Callan:1977pt}.
In this picture,
one assumes there is a scalar field with potential having two (or more) 
minima with at least one meta-stable minimum and an absolute minimum.
For simplicity, let us assume there is only one meta-stable minimum
(false vacuum) and only one absolute minimum (true vacuum).
If the universe is in the false vacuum in the beginning, 
it decays into the true vacuum via bubble nucleation.

In the original paper by Coleman, however,
the effect of gravity was not taken into account. Hence it was not so
clear when and how gravity could affect the process. Then
the false vacuum decay in the presence of gravity was studied in detail
by Coleman and De Luccia~\cite{Coleman:1980aw}.
 Although limited by the thin-wall assumption, they successfully clarified
several important issues, for example the fact that the size of the bubble
should always be smaller than the Hubble horizon scale of the false vacuum.

Subsequently, Hawking and Moss argued that a homogeneous instanton sitting
at a maximum of the potential dominates the false vacuum decay if
the barrier is too shallow~\cite{Hawking:1981fz}. Later it was more
clearly realized by Jensen and Steinhardt~\cite{Jensen:1983ac}
that the Coleman-De Luccia (CD) instanton merges to the Hawking-Moss (HM)
instanton as the potential gets flatter, and except for possible
oscillating solutions~\cite{Hackworth:2004xb}, there will be only
HM solutions when $|m^2|<4H^2$ where $m^2$ is the curvature of the potential
at its maximum and $H$ is the corresponding Hubble parameter,
$H^2=\kappa^2V_{\rm top}/3$ where $\kappa^2=8\pi G$.

Concerning the false vacuum decay from a flat potential,
a very interesting observation was made by Lee and Weinberg~\cite{Lee:1985uv}.
They considered an unstable potential and another potential of 
similar shape except that it has
a small barrier near the extremum of the potential.
As the unstable potential becomes extremely flat at
its extremum, the time scale for the field to roll
 down the potential classically would eventually exceed the time scale for
the quantum vacuum decay of the corresponding potential with a barrier.
If there were no tunneling solution for the unstable potential,
this would lead to an unacceptable conclusion that the unstable potential
is stabler than the corresponding potential with a barrier. 
Indeed such a solution describing tunneling without barriers had 
been found by Linde~\cite{Linde:1981zj} prior to 
Lee and Weinberg~\cite{Lee:1985uv}, who showed
that there exists $O(4)$-symmetric instantons 
for potentials without barriers, and rediscovered 
an exact analytical solution for $-\lambda\phi^4$ potential 
($\lambda>0$) found in~\cite{Linde:1981zj}.

However, neither Linde nor Lee and Weinberg included the effect of gravity.
It is then of great interest how gravity can modify the result. 
A pioneering study was done by Lee~\cite{Lee:1986saa}.
Recently an interesting oscillating solution was found for an inverted
quadratic potential by Lee et al.~\cite{Lee:2011ms} which wouldn't
be possible without gravity. Here, however, we focus on monotonic
solutions.

In the presence of gravity, the absolute value of the potential is
very important. If the false vacuum has a positive potential energy,
$V_F>0$, it is in the de Sitter phase with the Hubble parameter
$H=\sqrt{\kappa^2V_F/3}$. In this case, if we consider a flat potential
without barriers,
\begin{eqnarray}
V=V_F-\frac{\lambda}{4}\phi^4+\cdots,
\label{flatpot}
\end{eqnarray}
it is expected that the quantum decay of the unstable vacuum
will be mediated by the HM instanton rather than the 
Linde-Lee-Weinberg (LLW) instanton.

On the other hand, in the limit $V_F\to0$, 
the universe at $\phi=0$ will no longer be de Sitter but it is
simply a flat Minkowski space. In this limit, one would naturally
expect the LLW instanton to dominates the quantum vacuum decay
(of an unstable flat space). To clearly demonstrate this, and to
show explicitly when the LLW instanton takes over the HM instanton,
we need a solution for the LLW instanton in the presence of gravity.

In this paper, we present an exact, analytic solution of
the LLW instanton with gravity for a particular class of the potential
having the form (\ref{flatpot}) with $V_F=0$. 
For this solution we also obtain an analytical expression
for the value of the bounce action. Then we briefly discuss possible
implications of the solution to cosmology.

\section{Set-up}

We consider the Euclidean action for a scalar field $\phi$ with gravity:
\begin{eqnarray}
S_{E} =-\frac{1}{2\kappa^2}\int_{M} d^4x\sqrt{g}~R
- \frac{1}{\kappa^2}\int_{\partial M} d^{3}x \sqrt{h}~K
+\int_{M} d^4x \sqrt{g}\left[~
\frac{1}{2}g^{\mu\nu}\partial_\mu\phi\partial_\nu\phi
+V(\phi)
~\right]\,.
\label{basicaction}
\end{eqnarray}
The second term is the Gibbons-Hawking boundary term 
to make the variational principle consistent 
when the spacetime is non-compact~\cite{Gibbons:1976ue}. 

Assuming $O(4)$-symmetry, we consider the metric of the form,
\begin{eqnarray}
ds^2=a(z)^2\left(dz^2 + d\Omega^2_{3}\right)  \,, 
\end{eqnarray}
and assume $\phi=\phi(z)$. For an asymptotically Euclidean solution,
we have $a\propto e^{z}$ for $z\to\pm\infty$.
Under the $O(4)$-symmetry, the action reduces to
\begin{eqnarray}
S_E=2\pi^2\left[-\frac{3}{\kappa^2}\int dz
\left(\dot a^2+a^2\right)
+ \int dz\,a^{3}
\left(\frac{1}{2a}\dot\phi^2 + a\,V\right)\right]\,,
\label{action}
\end{eqnarray}
where the dot denotes a derivative with respect to $z$ ($\dot{~}=d/dz$).
The equations of motion are
\begin{eqnarray}
3\left(\frac{\ddot a}{a}-1\right)
=\kappa^2\left(-\frac{\dot\phi^2}{2}-2a^2V\right)\,,
\label{metric1}
\end{eqnarray}
and 
\begin{eqnarray}
\ddot\phi+2\frac{\dot a}{a}\dot\phi-a^2\frac{dV}{d\phi}=0  \, .
\label{sc1}
\end{eqnarray}
The Hamiltonian constraint, which is an integral of (\ref{metric1}) with
a specific integration constant, is
\begin{eqnarray}
3\left[\left(\frac{\dot a}{a}\right)^2-1\right]
=\kappa^2\left(\frac{1}{2}\dot\phi^2-a^2V\right)\, .
\label{hc1}
\end{eqnarray}

\section{analytical method}

We now construct an analytical LLW instanton solution with gravity
by extending the method developed in \cite{Kanno:2011vm}.
Namely, instead of giving the form of the potential first, we
consider the condition on the form of the scale factor for the
existence of a regular instanton solution and look for 
a function describing the scale factor that enables us to 
derive the potential as a function of the scalar field analytically.

Since we are interested in the case of asymptotically flat
solution, we may assume the form of the scale factor $a(z)$ as
\begin{eqnarray}
a(z)=e^z \ell\,g(\tanh z)\,,
\label{scalefactor}
\end{eqnarray}
where the function $g(\tanh z)$ is assumed to be regular at 
$z\to\pm\infty$, and $\ell$ is an arbitrary length scale.
This guarantees the asymptotic flatness of the metric.
For convenience, we introduce the variable $x$ by $x=\tanh z$.
In terms of $x$, we have
\begin{eqnarray}
e^z=\sqrt{\frac{1+x}{1-x}}\,,\quad
\frac{d}{dz}=(1-x^2)\frac{d}{dx}\,,
\quad
\frac{d}{dx}=\cosh^2z\frac{d}{dz}\,,
\end{eqnarray}
and the scale factor is expressed as
\begin{eqnarray}
a=\ell\sqrt{\frac{1+x}{1-x}}\,g(x)\,.
\end{eqnarray}
Then from Eqs.~(\ref{metric1}) and (\ref{hc1}), we can express
$\dot\phi^2$ and $V$ in terms of $\ddot a/a$ and $\dot a/a$, and hence
in terms of the function $g(x)$ and its derivatives.
The resulting expressions are
\begin{eqnarray}
&&\frac{\kappa^2}{2}\phi'{}^2
=\frac{2}{1-x}\frac{g'}{g}
+\left(\frac{g'}{g}\right)^2-\left(\frac{g'}{g}\right)'
\,,
\label{dphi}
\end{eqnarray}
and
\begin{eqnarray}
\kappa^2\ell^2V
= \frac{(1-x)^2}{g^2}
\left[(2x-4)\frac{g'}{g}-(1-x^2)\left\{\left(\frac{g'}{g}\right)'
+2\left(\frac{g'}{g}\right)^2\right\}
\right]\,,
\label{potential}
\end{eqnarray}
where the prime denotes an $x$-derivative (${}'=d/dx$).

Similarly, substituting (\ref{hc1}) into the action (\ref{action}),
it reduces to 
\begin{eqnarray}
S_E
& =& 4\pi^2 \int_{-\infty}^{\infty}dz
 \left[ a^4 V - \frac{3}{\kappa^2}\,a^2  \right]
\cr
&=&\frac{4\pi^2\ell^2}{\kappa^2}
\int_{-1}^{1}dx\frac{g^2}{1-x}
\left[-\frac{3}{1-x}
+(1+x)\left\{(2x-4)\frac{g'}{g}
-2(1-x^2)\left(\frac{g'}{g}\right)^2
-(1-x^2)\left(\frac{g'}{g}\right)'\right\}
\right]\,.
\label{actionwhc}
\end{eqnarray}
For an asymptotic flat solution, the above action diverges. However,
the difference between this action and the flat space action $S_0$
should be finite, where $S_0$ is given by
\begin{eqnarray}
S_0=\frac{4\pi^2\ell^2}{\kappa^2}
\int_{-1}^{1}dx\left[-3\frac{g_0^2}{(1-x)^2}\right]\,,
\end{eqnarray}
where $g_0=g(x=1)$.
The difference is the bounce action $B$ that appears in the
decay rate, $\Gamma\sim e^{-B}$, where
\begin{eqnarray}
B&\equiv& S_E-S_0
\cr
&=&\frac{4\pi^2\ell^2}{\kappa^2}
\int_{-1}^{1}dx\left[-3\frac{g^2-g_0^2}{(1-x)^2}
+g^2\frac{(1+x)}{1-x}\left\{(2x-4)\frac{g'}{g}
-2(1-x^2)\left(\frac{g'}{g}\right)^2
-(1-x^2)\left(\frac{g'}{g}\right)'\right\}
\right]\,.
\label{Bdef}
\end{eqnarray}
As one can easily see, if $g'(x)$ is non-zero at $x=1$ (ie, at infinity),
$B$ does not converge. Thus the boundary condition at infinity that
the solution be a legitimate asymptotically Euclidean instanton
is $g'(1)=0$. 

\section{exact solution}

As an example that satisfies the condition $g'(1)=0$,
we consider the case,
\begin{eqnarray}
g(x)=\frac{c}{\displaystyle c+\frac{(1-x)^2}{2}}
\quad(c>0)\,.
\end{eqnarray}
Inserting this into (\ref{dphi}), we find
\begin{eqnarray}
\frac{\kappa^2}{2} 
\left(\frac{d\phi}{dx}\right)^2
&=&\frac{6}{2c+(1-x)^2}\,.
\end{eqnarray}
This can be readily integrated to give
\begin{eqnarray}
\pm\frac{\kappa}{2}\phi=\sqrt{3}\,U\,,
\quad 1-x=\sqrt{2c}\,\sinh U\,.
\end{eqnarray}
Hence
\begin{eqnarray}
1-x=\pm\sqrt{2c}\,\sinh \frac{\kappa\phi}{\sqrt{12}}\,.
\label{xphi}
\end{eqnarray}

 From (\ref{potential}), the potential is given by
\begin{eqnarray}
\kappa^2\ell^2V
&=&-3\frac{(1-x)^4}{c^2}\left[c+2(1-x)-\frac{(1-x)^2}{2}\right]\,.
\end{eqnarray}
Inserting (\ref{xphi}) into the above, we obtain
a completely analytical potential 
(with $+$ sign for (\ref{xphi}) for definiteness),
\begin{eqnarray}
V=-\frac{12}{\kappa^2\ell^2}\sinh^4\frac{\kappa\phi}{\sqrt{12}}
\left[c\left(1-\sinh^2\frac{\kappa\phi}{\sqrt{12}}\right)
+2\sqrt{2c}\sinh\frac{\kappa\phi}{\sqrt{12}}\right]\,.
\end{eqnarray}
Setting $\lambda=c\kappa^2/(3\ell^2)$ and $\beta=2\sqrt{2/c}$,
we have
\begin{eqnarray}
V=-\frac{\lambda}{4}\left(\frac{12}{\kappa^2}\right)^2
\sinh^4\frac{\kappa\phi}{\sqrt{12}}
\left[1-\sinh^2\frac{\kappa\phi}{\sqrt{12}}
+\beta\sinh\frac{\kappa\phi}{\sqrt{12}}\right]\,.
\end{eqnarray}

\begin{center}
\begin{figure}
\includegraphics[width=15cm]{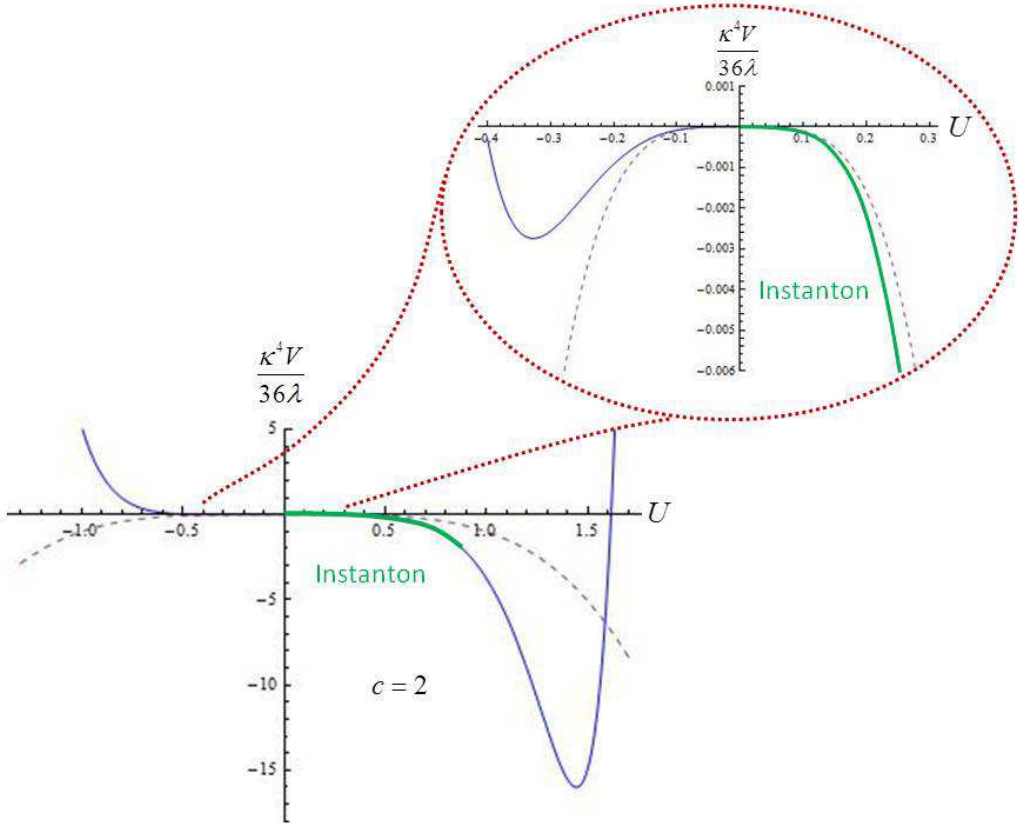}
\caption{\label{fig:LLWpot} 
A potential that allows an exact LLW instanton 
with gravity (drawn with the real line). 
The parameter is set to $c=2$ ($\beta=2$).
The horizontal axis is $U=\kappa\phi/\sqrt{12}$.
The thick line (in green) is the range the instanton runs.
The dotted curve shows the potential $V=-\lambda\phi^4/4$ for 
comparison. The upper right figure in the dotted circle
is a zoom-up of the potential near the maximum.
}
\end{figure}
\end{center}

The above is an exact, analytic Linde-Lee-Weinberg model {\it with gravity}.
For reference, the potential in the case of the parameter $c=2$ is shown
in Fig.~\ref{fig:LLWpot}. 
For $c\gg1$, the range of $\phi$ that the instanton
covers is small compared to the Planck scale, $\kappa|\phi|\ll1$. 
In this limit the above potential reduces to a simple $\phi^4$ potential,
\begin{eqnarray}
V=-\frac{\lambda}{4}\phi^4\Bigl(1+O(\beta\kappa\phi,\kappa^2\phi^2)\Bigr)\,.
\label{lambda}
\end{eqnarray}
The bounce action can be also evaluated analytically as
\begin{eqnarray}
B=4\pi^2\frac{\ell^2}{\kappa^2}
\left[
\frac{2 \sqrt{c}+3 \sqrt{2} (2+c) \arctan\sqrt{\frac{2}{c}}}
{4 \sqrt{c} (2+c)}\right]\,.
\end{eqnarray}
For large $c$, the above reduces to
\begin{eqnarray}
B=4\pi^2\frac{\ell^2}{\kappa^2}
\left[\frac{2}{c} - \frac{2}{c^2} + \frac{16}{5c^3} 
+ O\left(\frac{1}{c^4}\right)\right]\,.
\end{eqnarray}
In terms of the $\phi^4$ coupling constant $\lambda$ given
by (\ref{lambda}) this is re-expressed as
\begin{eqnarray}
B=\frac{8\pi^2}{3\lambda}
\left[1- \frac{\beta^2}{8}+\cdots\right]\,.
\end{eqnarray}
As expected the leading order term agrees with that of
the LLW instanton in flat space. We note that the value of
$\lambda$ is arbitrary: It can be of order unity or larger,
and hence the decay rate can become very large in principle.
However, if we assume $\lambda\lesssim1$, the limit $c\gg1$
($\beta\ll1$) implies $\ell\gg\kappa$, ie, the low energy limit.

One can also estimate an approximate size of the bubble.
Since the value of the scalar field at the center of the bubble is 
given by $\phi$ at $x=-1$, 
\begin{eqnarray}
\phi_{c}=\frac{\sqrt{12}}{\kappa}\sinh^{-1}\sqrt{\frac{2}{c}}
\approx\frac{2}{\ell}\sqrt{\frac{2}{\lambda}}\,,
\end{eqnarray}
for $c\gg1$, the radius at which the value of $\phi$ is half the
value at the center, $\phi\sim\phi_c/2$, is given approximately
by $a(x=0)\approx\ell$. Thus $\ell$ corresponds to the size of
the bubble.

\section{discussion}

We have found an exact, analytical LLW instanton in the presence of gravity.
The resulting bounce action reduces to that of the original LLW instanton
in flat space in the low energy limit,
\begin{eqnarray}
B_{LLW}=\frac{8\pi^2}{3\lambda}\,.
\end{eqnarray}

Recently tunneling between different vacua through the
false vacuum decay or even true vacuum decay~\cite{Lee:1987qc}
in the presence of gravity has attracted 
renewed interest in the context of the string theory 
landscape~\cite{Susskind:2003kw,Freivogel:2005vv}. 
In the landscape, there are an exponentially large number of
 different vacua with different values of the vacuum energy,
and tunneling between different vacua may occur frequently enough
to result in various interesting consequences which may be
 observationally tested, for example,
traces of bubble collisions~\cite{Czech:2010rg},
traces of false vacuum prior to inflation~\cite{Yamauchi:2011qq}, etc..
Here let us consider possible implications of LLW instantons
to cosmology.

Assuming there indeed exits a LLW-type scalar field,
let us first estimate the probability for a bubble to nucleate
within our past light-cone. The nucleation rate per unit volume
per unit time is estimated as
\begin{eqnarray}
\Gamma\sim\ell^{-4}e^{-B}\,.
\end{eqnarray}
Then the probability for a bubble to nucleate within our past light-cone
can be obtained by multiplying the above rate by the present
Hubble volume times the Hubble time,
\begin{eqnarray}
p\sim H_0^{-4}\Gamma\sim\frac{e^{-B}}{(H_0\ell)^4}
=\frac{9\lambda^2}{c^2}\frac{e^{-B}}{(\kappa H_0)^4}\,,
\end{eqnarray}
where we have used the relation, $\ell^{-2}=(3\lambda/c)\kappa^{-2}$.
Assuming $c\gg1$ and $\lambda\lesssim1$, the energy scale of the bubble
$\sim\ell^{-1}$ is much smaller than the Planck scale. In this case
we have $B=B_{LLW}=8\pi^2/(3\lambda)$ as we saw already.
Therefore, with an estimate that $\kappa H_0\sim 10^{-60}$,
we find
\begin{eqnarray}
p\sim \frac{9\lambda^2}{c^2}\times
10^{240}\exp\left[-\frac{8\pi^2}{3\lambda}\right]
\approx\frac{9\lambda^2}{c^2}
\exp\left[550-\frac{8\pi^2}{3\lambda}\right]\,.
\end{eqnarray}
For $c\gg1$, this will be exponentially smaller than unity for
$\lambda\lesssim0.01$.

In the above we have ignored the presence of dark energy.
To see its effect, 
let us consider the case where the false vacuum has an extremely 
small but non-vanishing vacuum energy with the Hubble parameter,
$H_F$, such that $H_F\ell\ll1$. If we apply this case
to the present universe, the value of $H_F$ is
approximately equal to the present Hubble constant $H_0$.
Then the universe appears almost flat over the
length scale of $\ell$, hence the use of the LLW instanton
which neglects the false vacuum energy is justified.
We further assume that there is a small,
very flat barrier and the Hubble parameter at the top of the barrier 
also satisfies $H_{top}\ell\ll1$.
In this case, there is a HM instanton and the bounce action 
will be given by
\begin{eqnarray}
B_{HM}=S_{HM}-S_F
=\frac{8\pi^2}{\kappa^2H_F^2}\left[1-\frac{H_F^2}{H_{top}^2}\right]\,.
\end{eqnarray}
If we compare this with the LLW bounce action, we find
\begin{eqnarray}
\frac{B_{LLW}}{B_{HM}}\approx\frac{\kappa^2H_F^2}{3\lambda}
\left[1-\frac{H_F^2}{H_{top}^2}\right]^{-1}
\approx\frac{H_F^2\ell^2}{c}\ll1\,,
\end{eqnarray}
unless the value of $H_{top}$ is exponentially close to $H_F$.
Thus the LLW instanton will dominate the false vacuum decay
in this situation. In other words, the presence of dark energy
does not alter the physics of tunneling.

The above consideration implies that if the potential near
the false vacuum is extremely flat and the minimum is below zero,
the vacuum decay will proceed much faster than what one would expect
from the value of the bounce action for the HM instanton even if
the potential at the false vacuum were positive. Boldly applying
 this to the present universe with the vacuum energy 
density $\sim3H_0^2/\kappa^2$,
the fact that there is no sign of the vacuum decay might
mean either that there is no minimum with a negative energy density
near the vacuum we live in or that the coupling constant
$\lambda$ is much smaller than unity, say $\lambda\lesssim0.01$.

\acknowledgements

This work is supported in part by the JSPS
Grants-in-Aid for Scientific Research (C) No.~22540274 and (A) No.22244030,
the Grant-in-Aid for Creative Scientific Research No.~19GS0219.
SK is supported by grant PHY-0855447 from the National Science Foundation.
SK would like to thank the Yukawa Institute for Theoretical Physics,
Kyoto University, for hospitality where this work was completed.


\begin{thebibliography}{99}
 
\bibitem{Coleman:1977py}
  S.~R.~Coleman,
  Phys.\ Rev.\  {\bf D15}, 2929-2936 (1977).

\bibitem{Callan:1977pt}
  C.~G.~Callan, Jr., S.~R.~Coleman,
  Phys.\ Rev.\  {\bf D16}, 1762-1768 (1977).
  
\bibitem{Coleman:1980aw}
  S.~R.~Coleman, F.~De Luccia,
  Phys.\ Rev.\  {\bf D21}, 3305 (1980).
 
\bibitem{Hawking:1981fz}
  S.~W.~Hawking, I.~G.~Moss,
  Phys.\ Lett.\  {\bf B110}, 35 (1982).

\bibitem{Jensen:1983ac}
  L.~G.~Jensen and P.~J.~Steinhardt,
  Nucl.\ Phys.\  B {\bf 237}, 176 (1984).

\bibitem{Hackworth:2004xb} 
  J.~C.~Hackworth and E.~J.~Weinberg,
  Phys.\ Rev.\ D {\bf 71}, 044014 (2005)
  [hep-th/0410142].

\bibitem{Lee:1985uv}
  K.~M.~Lee and E.~J.~Weinberg,
  Nucl.\ Phys.\  B {\bf 267}, 181 (1986).

\bibitem{Linde:1981zj} 
  A.~D.~Linde,
  Nucl.\ Phys.\ B {\bf 216}, 421 (1983)
  [Erratum-ibid.\ B {\bf 223}, 544 (1983)].


\bibitem{Lee:1986saa} 
  K.~-M.~Lee,
  Nucl.\ Phys.\ B {\bf 282}, 509 (1987).


\bibitem{Lee:2011ms} 
  B.~-H.~Lee, C.~H.~Lee, W.~Lee and C.~Oh,
  arXiv:1106.5865 [hep-th].

\bibitem{Gibbons:1976ue}
  G.~W.~Gibbons, S.~W.~Hawking,
  Phys.\ Rev.\  {\bf D15}, 2752-2756 (1977).
  
\bibitem{Kanno:2011vm}
  S.~Kanno and J.~Soda,
  arXiv:1111.0720 [hep-th].

\bibitem{Lee:1987qc}
  K.~M.~Lee and E.~J.~Weinberg,
  Phys.\ Rev.\  D {\bf 36}, 1088 (1987).

\bibitem{Susskind:2003kw}
  L.~Susskind,
  arXiv:hep-th/0302219.

\bibitem{Freivogel:2005vv} 
  B.~Freivogel, M.~Kleban, M.~Rodriguez Martinez and L.~Susskind,
  JHEP {\bf 0603}, 039 (2006)
  [hep-th/0505232].

\bibitem{Czech:2010rg} 
  B.~Czech, M.~Kleban, K.~Larjo, T.~S.~Levi and K.~Sigurdson,
  JCAP {\bf 1012}, 023 (2010)
  [arXiv:1006.0832 [astro-ph.CO]].
\\
  M.~Kleban, T.~S.~Levi and K.~Sigurdson,
  arXiv:1109.3473 [astro-ph.CO].

\bibitem{Yamauchi:2011qq} 
  D.~Yamauchi, A.~Linde, A.~Naruko, M.~Sasaki and T.~Tanaka,
  Phys.\ Rev.\ D {\bf 84}, 043513 (2011)
  [arXiv:1105.2674 [hep-th]].



\end{thebibliography}
\end{document}